\documentclass[showpacs,aps,prb,twocolumn]{revtex4}
\usepackage{epsfig}
\usepackage{multirow}

\def\dz{\delta_z}
\def\Ddij{\Delta d_{ij}}
\def\ei{\eta_i}

\begin{document}

\title{Coexistence of antiferrodistortive and ferroelectric 
distortions at the PbTiO$_3$ (001) surface }

\author{Claudia Bungaro}
\email{bungaro@physics.rutgers.edu}
\author{K. M.~Rabe}
\affiliation{Department of Physics and Astronomy, Rutgers University,
Piscataway, NJ 08854-8019, USA.}

\begin{abstract}
The c(2$\times$2) reconstruction of (001) PbTiO$_3$ 
surfaces is studied by means of first principles calculations for
paraelectric (non-polar) and ferroelectric ([001] polarized) films.
Analysis of the atomic displacements in the near-surface region shows 
how the surface modifies the antiferrodistortive (AFD)
instability and its interaction with ferroelectric (FE) distortions.
The effect of the surface is found to be termination dependent. The AFD
instability is suppressed at the TiO$_2$ termination while it is strongly
enhanced, relative to the bulk, at the PbO termination resulting in a
c(2x2) surface reconstruction which is in excellent agreement with 
experiments.
We find that, in contrast to bulk PbTiO$_3$,
in-plane ferroelectricity at the PbO termination does not suppress 
the AFD instability.
The AFD and the in-plane FE distortions are instead concurrently enhanced 
at the PbO termination. This leads to a novel surface
phase with coexisting FE and AFD distortions which is not found in
PbTiO$_3$ bulk.

\end{abstract}
\date{\today}
\pacs{  }

\maketitle

%%%%%%%%%%%%%%%%%%%%%%%%%%%%%%%%%%%%%%%%%%%%%%%%%%%%%%%%%%%%%%%%%%%%%%%%%%%%%%

%\marginparwidth 2.7in
%\marginparsep 0.5in
%\def\cb#1{\marginpar{\Red{\small CB: #1}}}

%% ******************************
%% ******************************
%\def\epsfig#1{}
%% ******************************
%% ******************************
%% This turns off actual insertion of figures, so that the
%% latex command runs faster.  Obviously it should be commmented
%% or removed in the final version, or whenever printing.

%\columnseprule 0pt

%\narrowtext

%%%%%%%%%%%%%%%%%%%%%%%%%%%%%%%%%%%%%%%%%%%%%%%%%%%%%%%%%%%%%%%%%%%%%%%%%%%%%%

\section{Introduction}

The ABO$_3$ perovskite structure exhibits several lattice
instabilities which lead to a rich variety of phase diagrams with
structures ranging from non-polar antiferrodistortive (AFD) to
ferroelectric (FE) and antiferroelectric (AFE).  In these systems,
polar and non-polar instabilities often compete and tend to suppress
one another. For instance, in SrTiO$_3$ it has been found that
ferroelectricity is enhanced if the AFD degrees of freedom are
artificially frozen out by setting the corresponding distortion to
zero.\cite{ZhongVanderbiltPRL95,Sai2000} Of these various instabilities, the
polar instabilities are the ones responsible for the technologically important
and fundamentally interesting properties, such as ferroelectricity and
large piezoelectric and dielectric responses.

Currently there is a great deal of effort being expended to develop
nanoscale devices. If nanoscale systems based on ferroelectricity are
to be feasible, then the polar instabilities must remain dominant down
to the nanoscale.  In systems of reduced dimensionality, such
as thin films or nanoparticles,
a substantial fraction of the material is found in
proximity to a surface, and surface effects may become important. The
surface can affect the material behavior by modifying the strength of
the various instabilities and their interactions leading to novel
phases not present in bulk systems.
 
Lead titanate (PbTiO$_3$) is the prototype of a large class of
Pb-based perovskites. The cubic perovskite structure of
PbTiO$_3$ possesses several branches of unstable phonons, including FE
instabilities which consist of zone-center polar TO modes and AFD
instabilities which consist of nonpolar zone-boundary modes involving
rotations of oxygen octahedral cages surrounding the Ti atoms.~\cite{Ghosez99}
In bulk PbTiO$_3$, the FE and AFD instabilities of the cubic structure
compete with each other and the FE lattice distortion suppresses the
AFD distortion so that AFD distortions do not participate in the
ground-state tetragonal FE structure.  The possible impact of the 
proximity of the surface on lattice instabilities and their mutual interactions
is thus of great importance in understanding the structure and
properties of PbTiO$_3$ surfaces and ultrathin films.

Previous theoretical studies on lead titanate have been confined to
the behavior of the polar FE instability at the (001)
surface.~\cite{Padilla97, Bernd99, Bernd01} 
These studies employed (1$\times$1)
surface periodicity, which by construction does not permit AFD
deformations. Proximity to the surface, however, could enhance the
strength of non-polar instabilities and modify their interaction with
the FE instability.  Indeed, an AFD c(2$\times$2) reconstruction has
recently been found at the (001) surface of PbTiO$_3$ using grazing
incidence x-ray scattering,\cite{ArgonnePRL02} suggesting that the AFD
instability is important at the surface.

In the following, we study the behavior of the AFD instability and its
possible competition with ferroelectricity at the PbTiO$_3$ (001)
surface using first-principles calculations.  Initially we study the
AFD distortion at the surface, using symmetry constraints to freeze
out polar distortions; next we relax this constraint and study the
competition between the AFD and in-plane FE distortions at the
PbO-terminated surface, which is the equilibrium termination for PbTiO$_3$
(001)\cite{Bernd99}.

In the absence of ferroelectricity, we find that the impact of the
surface is termination-dependent: at the PbO-terminated surface, the
strength of the AFD instability is enhanced relative to the bulk;
while at the TiO$_2$ terminated surface, the strength of the
instability is diminished. At the PbO-surface, in contrast to the
bulk, where the presence of ferroelectricity suppresses AFD, we find
that the two coexist and we observe a c(2$\times$2) reconstruction
that agrees very well with experiment.\cite{ArgonnePRL02}

The paper is organized as follows: section~\ref{sec:theory}
is devoted to the theoretical method and the technical details of our 
calculations. In section~\ref{sec:AFDbulk} we study the AFD instability 
in bulk cubic lead titanate. 
In section~\ref{sec:AFDsurface}  
we study the surface effect upon the AFD distortion in 
absence of FE distortion. The competition of AFD and in-plane FE distortions
in the near-surface region is studied in section~\ref{sec:AFD+FE}. 
The conclusions are summarized in section~\ref{sec:conclusion}.

\section{Theoretical method}
\label{sec:theory}

Our theoretical study is based on ab initio calculations that were
performed within density-functional theory (DFT) using the plane-wave
pseudopotential method as implemented in the PWSCF
package\cite{PWSCF}.  In this section we describe the technical
details of the calculations and the supercells used to study the
surface of PbTiO$_3$ (001) films.

\subsection{Computational details}

The exchange and correlation energy was given within the local density
approximation (LDA) using the parameterization of Perdew and
Zunger\cite{CAPZ}.
Ultrasoft pseudopotentials \cite{ultrasoft} were used to describe the
electron-ion interaction, treating as valence states the 3$s$, 3$p$,
3$d$, and 4$s$ states of Ti, the 4$s$, 4$p$, 4$d$, and 5$s$ states of
Zr, the 5$d$, 6$s$, and 6$p$ states of Pb, and the 2$s$ and 2$p$
states of O.  The wave-functions were expanded in plane waves with an
energy cutoff of 25 Ry.  A larger cutoff of 250 Ry was used for the
charge density, which includes the augmentation charges required by the
use of ultrasoft pseudopotentials.
The Brillouin zone (BZ) integration was performed using a (3,3,2)
Monkhorst-Pack mesh~\cite{kmp} in the Brillouin zone of the
c(2$\times$2) supercells.
Relaxations of atomic coordinates have been iterated until the forces
on the atoms were less than 0.001 Ry/a.u. (26 meV\AA$^{-1}$).

\subsection {Surface and slab geometries}

\begin{figure}[ht!]
\hbox{\hspace{2cm} \epsfig{figure=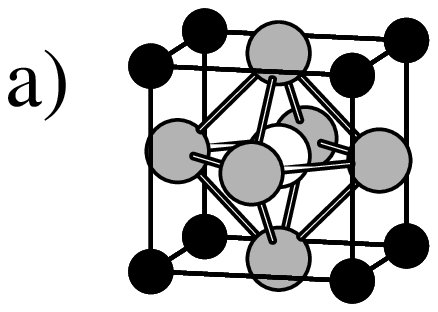,width=3cm,angle=-90}}
\hbox{
\epsfig{figure=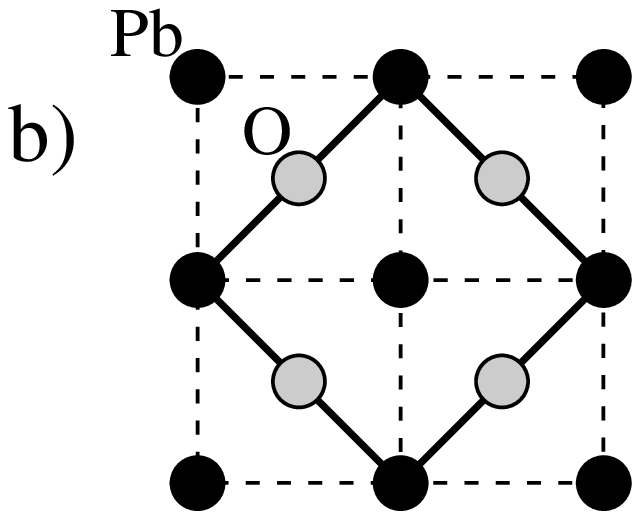,width=3cm,angle=-90}
\quad
\epsfig{figure=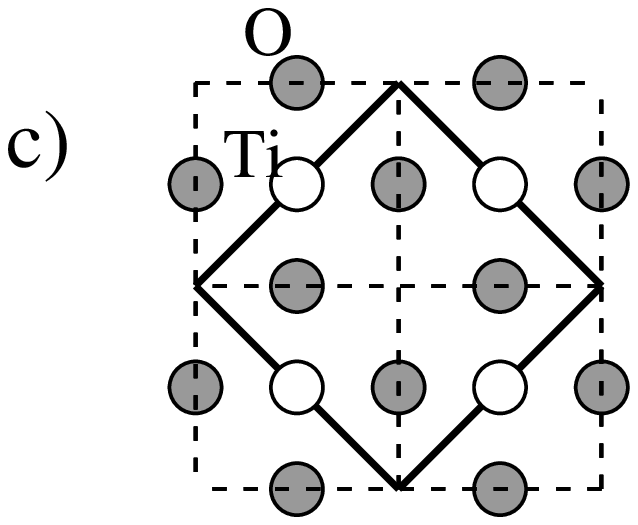,width=3cm,angle=-90}
}
\caption{(a) Crystal structure of
the ideal cubic perovskite unit cell of PbTiO$_3$. In (b) and (c) are
top views of the ideal PbO and TiO$_2$ (001) atomic layers;
the solid lines indicate the c(2$\times$2) surface unit cells.}
\label{fig:ideal-structure}
\end{figure}

In PbTiO$_3$, the stacking along [001] consists of alternating TiO$_2$
and PbO atomic layers (see Fig.~\ref{fig:ideal-structure}).  There are
two possible terminations for the non-polar (001) surface: the
PbO-terminated surface and the TiO$_2$-terminated surface.  We have
studied both types of surface termination.

The surfaces were modeled using a periodic slab geometry.
Specifically, we used symmetric slabs, terminated by two equivalent
surfaces, and separated by a vacuum region of approximately 14 \AA.
For the PbO-termination we used slabs of 11 atomic layers 
(7-layer slabs were used for direct comparison with previous calculations), and
for the TiO$_2$ termination we used slabs of 9 atomic layers. A
c(2$\times$2) in-plane supercell was used, containing 4 atoms in each
PbO layer and 6 atoms in each TiO$_2$ layer
(Fig.~\ref{fig:ideal-structure}).  The in-plane lattice parameter is
$a_0\sqrt{2}$, where $a_0$=7.37 a.u. is the theoretical lattice
parameter of cubic PbTiO$_3$ bulk.  This mimics the epitaxial
constraint imposed by a lattice-matched cubic substrate.

In order to investigate the behavior of the AFD distortion in the
absence of FE distortions, we studied non-polar (paraelectric)
c(2$\times$2) films. Specifically, we relaxed the atomic positions in the lowest symmetry 
space group containing inversion (P2/m),\cite{p2m} as this allows for the largest number of
degrees of freedom for the atomic displacements during the energy
minimization while constraining the spontaneous polarization to zero.
With this symmetry, all the atoms are allowed to move in
the $z$ (surface normal) direction and the oxygen atoms in the TiO$_2$
layers are also allowed to move in the $xy$ plane (parallel to the
surface). All the other atoms are fixed by symmetry to the ideal
perovskite coordinates in the $xy$ plane.  All the symmetry-allowed
displacements were fully relaxed to find the equilibrium structure
which minimizes the energy.

The competition between the AFD and FE distortions was studied using a
ferroelectric film with in-plane polarization.  This is achieved by
using a symmetric c(2$\times$2) slab with space group Pm.\cite{pm} The
mirror symmetry, M$_z$, prevents spontaneous polarization in
the $z$ direction and allows only an in-plane polarization to develop in
the film.

The case of perpendicular polarization is not considered in this work
because it requires special treatment of the electrical boundary
conditions and a separate discussion.\cite{dipole}

%%%%%%%%%%%%%%%%%%% FIGURE 1 %%%%%%%%%%%%%%%%%%%%%%%%%%%%%%%%%%%%%%%%

The atomic structure was optimized by relaxing the
symmetry-allowed atomic displacements.
We defined atomic displacements, $\delta_\alpha$, relative to the ideal 
perovskite structure.
Atomic relaxations normal to the surface, $\delta_z$, cause a rumpling of 
the atomic layers and a change in interlayer distances relative to the 
ideal cubic structure.  
The in-plane atomic relaxations, $\delta_\parallel$, chiefly
consist of rotations of the
oxygen squares surrounding the Ti atoms in the TiO$_2$ planes.

To analyze the optimized structures and compare with previous results
we computed the following parameters: the rumpling amplitude, $\ei$,
in each layer, $i$, the change in the interlayer distances relative to
the ideal cubic structure, $\Ddij$, and the rotation angle of the
oxygen squares in the TiO$_2$ planes, $\theta_i$.

The structural parameters are defined as follows.
For each atomic plane, $i$, we let $\overline\dz({\rm O}_i)$ and
$\overline\dz({\rm M}_i)$ be the average $z$ displacements of the
oxygen and metal atoms, respectively.  As in
Ref.~\onlinecite{Bernd99}, the interlayer distance $d_{ij}$ is given
by the difference of the averaged $\overline\dz$ displacements,
$[\overline\dz({\rm O}_i)+\overline\dz({\rm M}_i)]/2$, computed for
layer $i$ and $j$. The layer rumpling is given by
$\ei=[\overline\dz({\rm M}_i)-\overline\dz({\rm O}_i)]$.  A negative
rumpling, $\ei<0$, means that on average the metal atoms are deeper
inside the surface relative to the O atoms.
The rotation angles $\theta_i$ are zero for the (1$\times$1) surface, a value
different from zero indicating a c(2$\times$2) reconstruction.

\section{Antiferrodistortive instability in bulk cubic 
P\lowercase{b}T\lowercase{i}O$_3$}

\label{sec:AFDbulk}

To understand how the AFD instability is modified near the surface
we first study the AFD distortion in bulk PbTiO$_3$.
The rotation of oxygen octahedra is a common AFD instability in
perovskite oxides.  In bulk cubic PbTiO$_3$ it is associated with two
unstable zone-boundary phonons: the R$_{25}$ mode at the R point of the
Brillouin zone, {\bf q}=(111)$\pi/a$, and the M$_3$ mode at the M
point, {\bf q}=(110)$\pi/a$. The three-fold degenerate R$_{25}$ 
mode can be chosen to correspond to the
rotations, about the $\left<001\right>$ Cartesian axes,
of TiO$_6$ octahedra in opposite
directions from one cubic unit cell to the next. The singly-degenerate
M$_3$ mode is
similar to the R$_{25}$ mode, but neighboring planes of octahedra
along the rotation axis rotate in-phase instead of out-of-phase.

\begin{figure}[ht!]
\epsfig{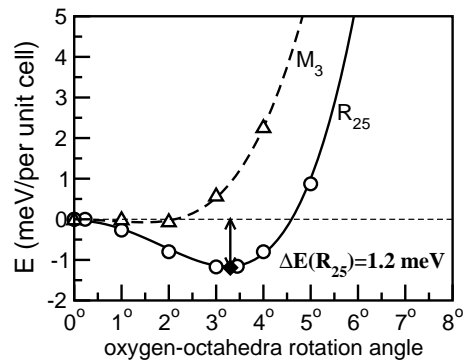}
\caption{Energetics of the AFD instability in bulk cubic PbTiO$_3$. Filled 
diamond indicates the minimum, $\theta = 3.3^o$ and $\Delta E = 1.2$ meV 
per bulk unit cell, to be compared with the ferroelectric well depth, 
$\Delta E=50$ meV per bulk unit cell. Triangles and circles  
are calculations for the M$_3$ and R$_{25}$ distortions, respectively. 
The solid line is a polynomial interpolation.}
\label{bulk_R25_fig}
\end{figure}

We considered distorted structures obtained by separately ``freezing in" 
R$_{25}$ and M$_3$ modes in cubic PbTiO$_3$ at the theoretical cubic
lattice constant ($a_0$=7.37 a.u.).  In Fig.~\ref{bulk_R25_fig} we show
the computed values of the total energy per 5-atom unit cell as a
function of the rotation angle. The energy is lowered by the AFD distortion, 
as expected, and the R$_{25}$ distortion with rotation along a Cartesian
axis is more favorable than
the related M$_3$ distortion. 
The computed equilibrium value of the
octahedral rotation angle is $\theta_{\rm bulk}=3.3^o$, and the energy
gain associated with the AFD instability in the bulk is $\Delta E^{\rm
bulk}_{\rm AFD}$ = 1.2 meV per bulk unit cell (well depth in
Fig.~\ref{bulk_R25_fig}).

The small energy gain indicates that cubic PbTiO$_3$, in the absence
of any FE distortion, displays only a weak AFD instability.  In
particular, it is much weaker than the ferroelectric instability, which has a
well depth of about 15 meV for the cubic undistorted unit cell (and
about 50 meV per unit cell if coupling with the tetragonal distortion
is included).

The AFD instability is suppressed by the FE distortion. In fact the
ground-state ferroelectric tetragonal structure with 5 atoms per unit cell
has no unstable phonons at any wavevector\cite{Garcia1996}.  
The ferroelectric structure results from the
coupling of a FE polar distortion (TO mode) and a tetragonal strain.
We checked, therefore, if the suppression of the AFD instability can be
caused by the FE distortion alone.  To do this we first relaxed the
ferroelectric distortion without coupling to tetragonal strain by
keeping the unit cell fixed to the ideal cubic structure and allowing
internal atomic displacements along the (001) direction.  As a result
a spontaneous polarization develops along the (001) direction.  We
then computed the phonons for this ``cubic polarized'' structure (C-FE)
and found no
unstable modes at the R point of the Brillouin zone.  We conclude that
the FE distortion suppresses the AFD instability even in absence of
tetragonal strain.

%==================== FIGURE 2 =====================================
%===================================================================

\section{P\lowercase{b}T\lowercase{i}O$_3$ (001) surfaces in the absence of 
Ferroelectric distortions} 

\label{sec:AFDsurface}

To understand the behavior of non-polar distortions near the surface,
we study the atomic structure of the non-polar (paraelectric) films.
We studied both the PbO and TiO$_2$ terminations with c(2$\times$2)
in-plane translational symmetry. Our primary focus in this section
will be on
characterizing the c(2$\times$2) reconstruction and its dependence
on termination.

\subsection{PbO termination}

\begin{figure}[ht!]
\centerline{
\vbox{
\hbox{
\epsfig{figure=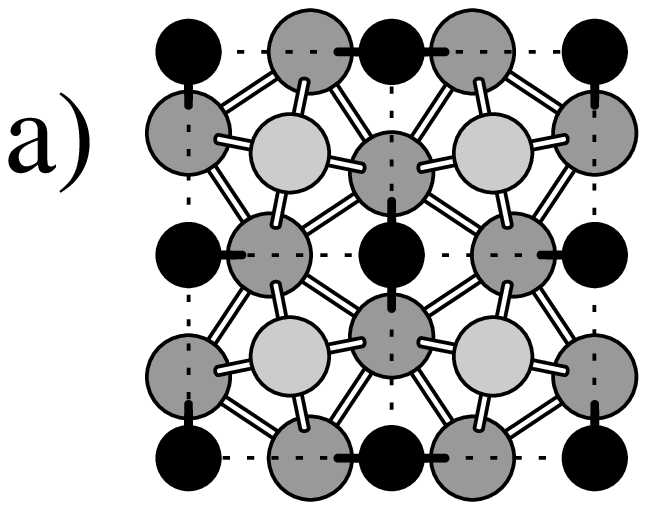,height=4.1cm,angle=-90}
}
\hbox{
\epsfig{figure=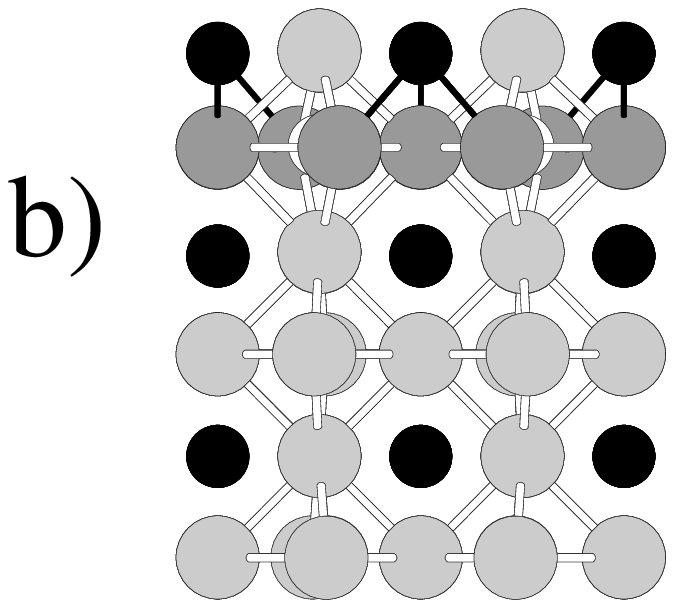,height=4cm,angle=-90}
}
}
}
%\centerline{ top (1$\times$1) \hspace{3cm} top c(2$\times$2) }
\caption{Optimized atomic structure of the c(2$\times$2) non-polar (001) 
PbTiO$_3$ film with PbO surface termination.  (a) Top view; only the
first top two layers are shown.  (b) Side view; only the top half of
the 11-layer slab is shown.  The oxygen atoms in the second layer are
shown in a darker shade of gray.  The c(2$\times$2) AFD
reconstruction, consisting of rotation of the Ti-centered oxygen
squares in the TiO$_2$ subsurface layer, is clearly visible.
Only Pb-O bonds shorter than 2.6 \AA\ are shown (in black); 
they are 2.39 \AA\ long.}
\label{fig:PbOnonpolar}
\end{figure}

Results for the atomic displacement and structural parameters of the optimized
c(2$\times$2) surface structure, shown in Fig. \ref{fig:PbOnonpolar}, are presented
and compared with
those for the (1$\times$1) surface in Tables~\ref{tab:PbOpositions}
and~\ref{tab:PbOstructure}.

The results for the (1$\times$1) surface are in excellent agreement
with a previous calculation~\cite{Bernd99}, given in Table
~\ref{tab:PbOstructure}. Small differences are chiefly due
to the different slab thickness used in the two calculations 
(we obtained almost identical results using a 7-layer slab as 
in Ref.\onlinecite{Bernd99}). We used
a thicker 11-layer slab to study the behavior of lattice 
distortions in the near-surface region and their decay to  
bulk-like behavior in the central layers of the slab.  
The change in interlayer distance displays an oscillating
behavior: the first interlayer distance, $d_{12}$, is substantially
contracted, whereas the second, $d_{23}$, is expanded and the third,
$d_{34}$, is again contracted.  The most noticeable features is the large
inward displacement of the surface Pb atom relative to the surface O atom.
This accounts for the large first interlayer contraction, $\Delta
d_{12}$, and first layer rumpling, $\eta_1$.
\begin{table} [ht!]
\caption{Atomic displacements, relative to the ideal cubic perovskite 
structure,
for the PbO-terminated (001) surface of non-polar PbTiO$_3$ films.
The relaxations perpendicular ($\dz$) and parallel ($\delta_\parallel$) 
to the surface
are given as a percentage of the cubic lattice parameter $a_0$.}
\begin{ruledtabular}
\begin{tabular}{l c   c c | c }
      &      & \multicolumn{2}{c|}{c(2$\times$2)}& (1$\times$1)    \\
layer & atom &  $\dz$   & $|\delta_\parallel|$ & $\dz$      \\
\hline
1     & Pb   & $-$1.83  &               & $-$4.34     \\
      & O    & $-$0.39  &               & $-$0.41     \\
2     & Ti   &   +2.79  &               &   +2.67     \\
      & O    &   +1.85  &   10.03       &   +1.26     \\
3     & Pb   & $-$1.57  &               & $-$2.08     \\
      & O    &   +0.41  &               & $-$0.35     \\
4     & Ti   &   +0.51  &               &   +0.64     \\
      & O    &   +0.09  &   2.52        &   +0.15     \\
5     & Pb   & $-$0.25  &               & $-$0.56     \\
      & O    & $-$0.02  &               & $-$0.11     \\
6     & Ti   &   0      &               &     0       \\
      & O    &   0      &   3.43        &     0       \\
\end{tabular}
\end{ruledtabular}
\label{tab:PbOpositions}
\end{table}
%\squeezetable

\begin{table} [ht!]
\caption{Structural parameters for the PbO-terminated (001) surface of 
non-polar PbTiO$_3$ films.
Change in the interlayer distance, $\Delta d_{ij}$, and 
layer rumpling, $\eta_i$, in percent of the lattice constant $a_{\rm bulk}$.
Rotation angle, $\theta_i$, of the oxygen squares in the TiO$_2$ layers. 
Relaxation energy relative to the ideal (1$\times$1) surface, 
$\Delta$E$_{\rm relax}$, in eV per (1$\times$1) surface unit cell.
}
\begin{ruledtabular}
\begin{tabular}{l c c c }
      & \multicolumn{2}{c}{this work} &Meyer et al.$^a$  \\
                & c(2$\times$2)   &  (1$\times$1) &  (1$\times$1)             \\
\hline
 $\Delta d_{12}$&  $-$3.4  & $-$4.3 & $-$4.2              \\
 $\Delta d_{23}$&    +2.9  &   +3.2 &   +2.6  \\
 $\Delta d_{34}$&  $-$0.9  & $-$1.6 & $-$0.8  \\
 $\Delta d_{45}$&    +0.4  &   +0.7 &         \\
 $\Delta d_{56}$&  $-$0.1  & $-$0.3 &         \\
\\

 $\eta_1$       &  $-$1.4   & $-$3.9 & $-$3.9  \\
 $\eta_2$       &    +0.9   &   +1.4 &   +1.2  \\
 $\eta_3$       &  $-$2.0   & $-$1.7 &  $-$1.2     \\
 $\eta_4$       &    +0.4   &   +0.5 &             \\
 $\eta_5$       &  $-$0.2   & $-$0.4 &             \\
\\

 $\theta_2$     & \ 11.4$^o$&     0  &    0        \\
 $\theta_4$     &  $-$2.9$^o$  &  0  &    0        \\
 $\theta_6$     &  \ \ 3.9$^o$ &  0  &             \\
($\theta_{\rm bulk}$)    & (3.3$^o$) &   &     \\
\\

$\Delta$E$_{\rm relax}$       
                & $-$0.29\ \ & $-$0.18&   \\ 
\end{tabular}
\end{ruledtabular}
\vspace{0.2cm}
$^a$Ref.\onlinecite{Bernd99}\\
\label{tab:PbOstructure}
\end{table}

The optimized atomic structure for the PbO termination of the
non-polar c(2$\times$2) PbTiO$_3$ (001) film is shown in
Fig.~\ref{fig:PbOnonpolar}, displaying the large c(2$\times$2)
reconstruction occurring in the TiO$_2$ subsurface layer.  The
reconstruction consists of alternating clockwise and anti-clockwise
rotations of the Ti-centered oxygen squares in the TiO$_2$ layer.  The
in-plane oxygen displacements are 10\% of the bulk lattice parameter,
corresponding to a rotation angle of $\theta_1$=$11^o$. This atomic
distortion is essentially an AFD rotation of the surface
oxygen-octahedra about the $z$ axis; particularly noteworthy is the
substantial enhancement of the AFD distortion at the PbO-terminated
surface. Neighboring planes of octahedra along the rotation axis
rotate out of phase, as in the dominant AFD bulk instability
(R$_{25}$ mode). 
Deep inside the film, the AFD distortion recovers its bulk
value in the absence of a ferroelectric distortion.
The relaxation energy, relative to the ideal (1$\times$1) surface, is
 $-0.29$ and $-0.18$ eV per (1$\times$1) surface unit cell for the
 c(2$\times$2) and (1$\times$1) surfaces, respectively.  The
 c(2$\times$2) reconstructed surface is much more favorable than the relaxed
 (1$\times$1) surface, the difference in energy being 0.11 eV per (1$\times$1) 
surface unit cell.
%%%%%%
It is interesting to notice that the symmetry of the relaxed structure
(space group P4/nmm) is higher than the minimal symmetry constraint imposed 
during energy minimization (space group P2/m).
The four oxygen atoms in each TiO$_2$ layer are equivalent by symmetry in 
the relaxed structure. 
%%%%%%
  
%==================== FIGURE 3 =====================================
%===================================================================

The main differences between the c(2$\times$2) and (1$\times$1)
structures are (i) the AFD rotation of the TiO$_4$-squares, chiefly in the
second layer, (ii) a much smaller inward displacement of the surface
Pb atoms, $\delta_z(Pb_1)=-1.83$ \% of $a_0$, (iii) and, consequently,
a substantial reduction of the first layer rumpling ($\eta_1$) and the
first interlayer contraction ($\Delta d_{12}$).

%%%%%%
These results can be understood in terms of the particular chemistry
of lead. In the ideal cubic perovskite bulk, Pb is in the center of an
O$_{12}$ cage and the twelve equivalent Pb-O bonds are 2.76~\AA\
long. The tendency of lead to move off-center and form stronger 
covalent PbO bonds, as short as 2.59\AA, has
been demonstrated to be an important factor in
ferroelectricity in PbTiO$_3$.\cite{PbANDferroel,bondNote} 
% In PbO the shortest Pb-O bond is 2.3~\AA\ long.
Formation of shorter bonds is favored in the broken symmetry
environment at the surface and, as we discuss next, can be 
regarded as the driving force
for the c(2$\times$2) AFD reconstruction occurring at the PbO-terminated surface.  

In Table~\ref{tab:bonds} we show the values of the interlayer bondlengths 
computed for the c(2$\times$2) and (1$\times$1) surfaces. Each
Pb atom forms four bonds with the in-plane oxygen atoms, four
interlayer bonds with O atoms in the plane above, and four interlayer
bonds with O atoms in the plane below; the surface Pb atom is
undercoordinated and has eight bonds (four in-plane bonds and four
interlayer bonds with the plane below).  In
the non-polar slab, in-plane Pb off-centering is not allowed by
symmetry. Thus, the four equivalent in-plane Pb-O bonds can only change by
bending; their lengths are essentially fixed to 2.76~\AA~and are
not included in Table~\ref{tab:bonds}. In contrast, the
interlayer Pb-O bonds can stretch or contract substantially.  When a
(1$\times$1) periodicity is enforced, 
%all four interlayer bonds, Pb$_i$$-$O$_j$, have the same length. 
the only possibility to shorten
the Pb-O bond, as found in Ref.~\onlinecite{Bernd99}, is by a large
downward displacement of the surface Pb atom, resulting in 
bondlengths (2.61 ~\AA\ long) comparable to those in tetragonal FE bulk.  
At the c(2$\times$2)
surface the AFD distortion offers an alternative way, evidently more
energetically favorable, to form fewer but shorter Pb-O bonds. 
Each surface Pb atom has
four oxygen neighbors in the subsurface TiO$_2$ layer. 
Rotation of the
TiO$_4$-squares brings two of these much closer, resulting in even 
stronger bonds 
(2.39 ~\AA\ long) only a bit longer than those
in bulk PbO (2.32 ~\AA\ long, experimental value~\cite{PbObond}). 
The rotation also shortens significantly two of the bondlengths for the 
Pb$_3$ atom
to 2.60~\AA~, comparable to those in tetragonal FE bulk.
Since the rotation provides a more effective alternative mechanism for
shortening Pb-O bonds, the inward displacement of surface Pb atoms (and the
related rumpling $\eta_1$ and interlayer contraction $\Delta d_{12}$)
is substantially reduced with respect to the (1$\times$1) surface.
The average bond length for each Pb atom does not change
much between the c(2$\times$2) and (1$\times$1) surfaces and it is not
much different from the undistorted bond length.

As a further verification of the particular role of lead in the
enhancement of the AFD distortion at the surface, we have relaxed the
non-polar (001) surface of BaTiO$_3$. The Ba-O bond has a more ionic
character and is therefore a good contrasting example. We find no AFD
distortion at the BaO surface. This confirms that the formation of
stronger covalent bonds is the origin of the AFD reconstruction at the
PbO-terminated (001) surface of PbTiO$_3$.

Finally we notice that, contrary to the Pb-O bonds,
the interlayer Ti-O bonds are almost the same at the c(2$\times$2) and
(1$\times$1) surfaces, and their values reflect the alternating
contraction and expansion of the interlayer distances.

\begin{table} [ht!]
\caption{Computed values for the Pb-O and Ti-O interlayer bond lengths
(in \AA\ ) at the PbO terminated c(2$\times$2) and (1$\times$1)
surfaces of the non-polar (001) PbTiO$_3$ film. For comparison, the 
in-plane Pb-O bond
length is 2.76~\AA. Average bond lengths Pb$_i$-O for each
Pb$_i$ atom are given in the ``average'' columns. 
The labels $i$ and $j$ indicate the atomic planes, with $i$,$j$=1 being the 
first surface layer. Shorter
Pb-O bonds are highlighted in bold.}
\begin {ruledtabular}
\renewcommand{\multirowsetup}{\centering}
\newlength{\LL} \settowidth{\LL}{2.78}
\begin{tabular}{c c c c  c c }
 & \multicolumn{3}{c }{c(2$\times$2) surface } &
                                \multicolumn{2}{c}{(1$\times$1) surface} \\ 
Pb$_i$$-$O$_j$  & short       & long & average&            & average \\ 
\hline
Pb$_1$$-$O$_2$  & {\bf 2.39}  & 2.96 & 2.72 & {\bf 2.61} & 2.69  \\
%\multicolumn{4}{c|}{ }\\
Pb$_3$$-$O$_2$  & {\bf 2.60}  & 3.13 &\multirow{2}{\LL}{2.78}  
                                            & 2.85       &\multirow{2}{\LL}{2.77}\\
Pb$_3$$-$O$_4$  &      2.64   & 2.78 &      & 2.70       &       \\
%\multicolumn{4}{c|}{ }\\
Pb$_5$$-$O$_4$  & 2.70        & 2.84 &\multirow{2}{\LL}{2.76}
                                            & 2.78       &\multirow{2}{\LL}{2.76}\\
Pb$_5$$-$O$_6$  & 2.66        & 2.85 &      & 2.74       &       \\
\multicolumn{4}{c }{ }\\
Ti$_i$$-$O$_j$ &\multicolumn{3}{c }{ }  \\
\hline
Ti$_2$$-$O$_1$ & 1.83 & \multicolumn{2}{c}{ }&1.83\\
Ti$_2$$-$O$_3$ & 2.04 & \multicolumn{2}{c}{ }&2.07\\
Ti$_4$$-$O$_3$ & 1.95 & \multicolumn{2}{c}{ }&1.91\\
Ti$_4$$-$O$_5$ & 1.97 & \multicolumn{2}{c}{ }&1.98\\
Ti$_6$$-$O$_5$ & 1.95 & \multicolumn{2}{c}{ }&1.95\\
\end{tabular}
\end{ruledtabular}
\label{tab:bonds}
\end{table}

\subsection{TiO$_2$ termination}

The computed structural parameters for the TiO$_2$ termination of the
non-polar film are summarized in Table~\ref{tab:TiO2structure}.  
As we discuss in detail below, the
large AFD surface reconstruction observed at the PbO termination does
not occur at the TiO$_2$ termination.

Our results for the (1$\times$1) TiO$_2$ surface are in very good agreement
with previous calculations.
The structure of the optimized c(2$\times$2) supercell is little changed from
that obtained for the (1$\times$1) slab.
The only difference is a small AFD rotation of the oxygen octahedra in
the inner layers, which is due to the
AFD instability of bulk PbTiO$_3$ in absence of FE distortion.
The AFD rotation in the first surface layer is essentially zero
($\theta_1=0.1^o$) indicating that, in contrast to the PbO termination,
the AFD instability is weakened at the TiO$_2$ termination.
The optimized c(2$\times$2) and (1$\times$1) structures have 
the same energy within numerical accuracy (they differ by less than
10$^{-6}$ Ry per atom).

Mayer et al.~\cite{Bernd99} have shown that in thermodynamic equilibrium ,
the (1$\times$1) PbO-termination is always more stable  than the (1$\times$1)
TiO$_2$-termination. Our results show that the TiO$_2$-termination
does not reconstruct while the PbO-termination is further stabilized
by the AFD c(2$\times$2) reconstruction. Therefore, the PbO termination
remains the most favorable.

The computed values for the Pb-O and Ti-O interlayer bond lengths are
shown in Table~\ref{tab:TiO2bonds}. The short and long Pb-O bonds are
very similar due to the very small AFD rotation angles, and they are
essentially equal at the surface where the rotation angle is
zero within the accuracy of the calculation. The average bond length for each Pb atom is equal to
the Pb-O bond length in the ideal perovskite structure (2.76~\AA\ ). The
interlayer Ti-O bond lengths are very similar to those at the
c(2$\times$2) and (1$\times$1) PbO terminations (see Table~\ref{tab:bonds}) 
indicating that they are independent of both termination and reconstruction.

\begin{table} [ht!]
\caption{Structural parameters for the TiO$_2$-terminated (001) surface of
paraelectric PbTiO$_3$ films.
Change in the interlayer distance, $\Delta d_{ij}$, and
layer rumpling, $\eta_i$, in percent of the lattice constant $a_{\rm bulk}$.
Rotation angle, $\theta_i$, of the oxygen squares in the TiO$_2$ layers.
Relaxation energy relative to the ideal (1$\times$1) surface, 
$\Delta$E$_{\rm relax}$, in
eV per (1$\times$1) surface unit cell.
}
\begin{ruledtabular}
\begin{tabular}{l c c c }
        & \multicolumn{2}{c}{this work} & Meyer et al.$^a$ \\
        &c(2$\times$2)   &   (1$\times$1)   &  (1$\times$1)  \\
\hline
 $\Delta d_{12}$& $-$4.3  & $-$4.3    &     $-$4.4       \\
 $\Delta d_{23}$&   +3.4  &   +3.3    &       +3.1       \\
 $\Delta d_{34}$& $-$1.2  & $-$1.3    &     $-$0.6       \\
 $\Delta d_{45}$&   +0.6  &   +0.5    &                  \\
\\

 $\eta_1$       & $-$3.2  & $-$3.2    &     $-$3.1       \\
 $\eta_2$       &   +4.3  &   +4.3    &       +4.1       \\
 $\eta_3$       & $-$1.0  & $-$0.9    &     $-$0.7       \\
 $\eta_4$       &   +0.9  &   +0.9    &                  \\
 $\eta_5$       & \ \ 0.0 &\ \ 0.0    &                  \\
\\

 $\theta_1$     & $-$0.1$^o$  &  0  &  0   \\
 $\theta_3$     & $-$2.1$^o$  &  0  &  0   \\
 $\theta_5$     & \ \ 2.8$^o$ &  0  &       \\
($\theta_{\rm bulk}$)    & (3.3$^o$) &   &  \\
\\

$\Delta$E$_{\rm relax}$ 
                & $-$0.26 &$-$0.26   &                  \\
\end{tabular}
\end{ruledtabular}
\label{tab:TiO2structure}
\end{table}

\begin{table} [ht!]
\caption{Pb-O and Ti-O interlayer bond length in \AA\ computed for the
TiO$_2$ termination of the c(2$\times$2) PbTiO$_3$ (001) surface. 
Average Pb-O bond lengths are also reported. We use the same notation as in
Table~\ref{tab:bonds}. The
in-plane Pb-O bond lengths are 2.76~\AA\ long. }
\begin {ruledtabular}
\renewcommand{\multirowsetup}{\centering}
\newlength{\Lbis} \settowidth{\Lbis}{2.76}
\begin{tabular}{c c c c | c c}
Pb$_i$$-$O$_j$  & short & long & average&
                            \multicolumn{2}{c}{Ti$_i$$-$O$_j$}\\
%                &       &      & Pb$_i$$-$O \\
\hline
Pb$_2$$-$O$_1$  & 2.62  & 2.63 &\multirow{2}{\Lbis}{2.76}
                                      &Ti$_1$$-$O$_2$& 1.80 \\
Pb$_2$$-$O$_3$  & 2.85  & 2.95 &      &Ti$_3$$-$O$_2$& 2.02 \\
Pb$_4$$-$O$_3$  & 2.67  & 2.78 &\multirow{2}{\Lbis}{2.76}
                                      &Ti$_3$$-$O$_4$& 1.90 \\
Pb$_4$$-$O$_5$  & 2.72  & 2.86 &      &Ti$_5$$-$O$_4$& 1.95 \\
\end{tabular}
\end{ruledtabular}
\label{tab:TiO2bonds}
\end{table}

\subsection{Surface effect on the strength of the AFD instability}

The energy gain associated with the lattice deformation is a measure
of the strength of the instability, and can
be used to compare the strength of the AFD instability at the
PbO-surface with the one in the bulk. 

The energy gain associated with the AFD deformation in a slab can be
defined as the difference between the energy of the relaxed
(1$\times$1) non-polar slab, in which by symmetry there is no octahedron rotation,
and the energy of the relaxed c(2$\times$2) non-polar slab.  This
energy difference can be divided into a {\it bulk} and a {\it surface}
contribution as follows:
\begin{equation}
%the factor 2 is because 1$\times$1 unit cell is half c(2$\times$2) unit cell
2 E^{\rm (1 \times 1) slab } - E^{\rm c(2 \times 2) slab} = \\
N \Delta E^{\rm bulk}_{\rm AFD} + 4\Delta E^{\rm surf.}_{\rm AFD}
\label{eq:energy}
\end{equation}
The bulk contribution, $N \Delta E^{\rm bulk}_{\rm AFD}$, is the
energy gain associated with the bulk-like rotation of the inner
``non-surface'' octahedra, with N being the number of such bulk-like
octahedra in the c(2$\times$2) unit cell (N=6 for the 11 layer slab).
The surface contribution, $4\Delta E^{\rm surf.}_{\rm AFD}$, is the
energy gain associated with the rotation of the surface octahedra.  The
factor of four arises because in the unit cell of a c(2$\times$2) slab there
are four surface octahedra, two at each surface of the slab.
Using Eq.~\ref{eq:energy}, and the computed values for the three
energies: $E^{\rm (1 \times 1) slab }$, $E^{\rm c(2 \times 2) slab}$,
and $\Delta E^{\rm bulk}_{\rm AFD}$, we find that the energy
associated with the surface AFD distortion at the PbO-termination is
$\Delta E^{\rm surf.}_{\rm AFD}$= 109 meV per (1$\times$1) surface
unit cell. 

By comparing this result with the AFD and FE well depths in the bulk 
(1.2 meV and 50 meV, respectively) we find that the 
energy gain associated with the surface AFD rotation
is two orders of magnitude larger at the PbO-terminated surface than
in the bulk, and about twice the bulk ferroelectric well depth.

\section{In-plane polarized P\lowercase{b}T\lowercase{i}O$_3$ (001) films} 
\label{sec:AFD+FE}

In this section we present our results for the c(2$\times$2) PbTiO$_3$
(001) film with spontaneous polarization parallel to the surface,
{\bf P}=P(100). In particular we are interested in studying the
interaction between AFD and in-plane FE distortions in proximity of
the (001) surface.  We consider only the PbO termination, which is the
most energetically favorable and, as we have shown in the previous section, displays
a strong surface AFD instability in absence of FE distortions.

\begin{figure}[!]
\centerline{
\vbox{
\hbox{
\epsfig{figure=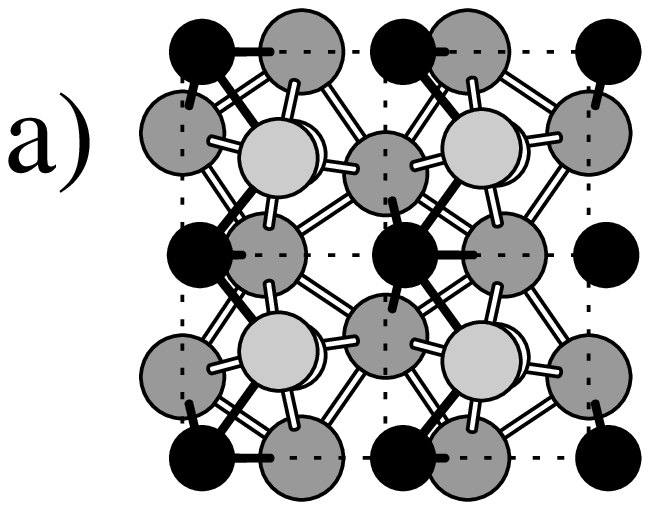,height=4cm,angle=-90}
}
\vspace{0.3cm}
\hbox{
\epsfig{figure=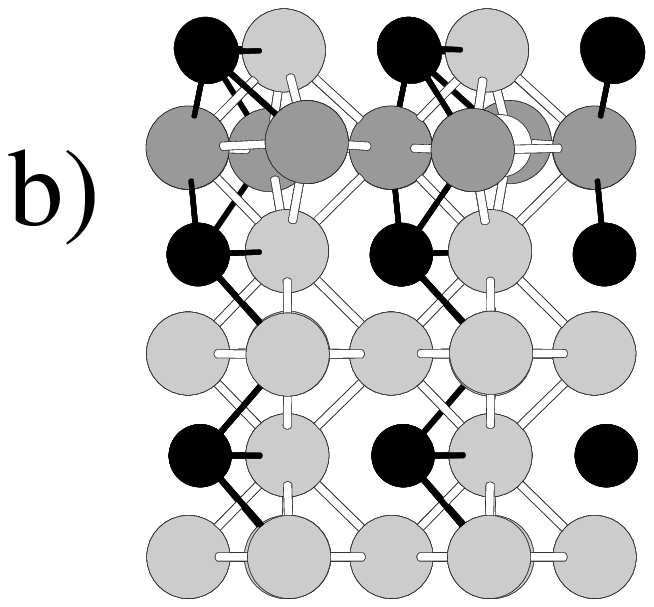,height=4cm,angle=-90}
}
}
}
\caption{Optimized atomic structure of the in-plane polarized
c(2$\times$2) (001) PbTiO$_3$ film with PbO surface termination.
(a) Top view; only the first top two layers are shown.  (b) Side
view; only the top half of the 11-layer slab is shown.  The oxygen
atoms in the second layer are shown in a darker shade of gray. The
in-plane Pb off-centering as well as the c(2$\times$2) AFD O-rotation
in the TiO$_2$ subsurface layer, are clearly visible. Pb-O bonds shorter than
2.6 \AA\ are shown in black.}
\label{fig:inplane}
\end{figure}

The optimized atomic structure of the c(2$\times$2) in-plane polarized
film is shown in Fig.~\ref{fig:inplane}. Clearly visible are: (i) the
in-plane Pb off-centering, present throughout the film and enhanced at
the surface, and (ii) the large AFD TiO$_4$ rotation in the TiO$_2$
subsurface layer. Thus, contrary to the bulk where FE distortions
suppress the AFD rotation, FE and AFD distortions coexist at the PbO
surface.
%====================== FIGURE 4 ========================================
%========================================================================

\begin{figure*}
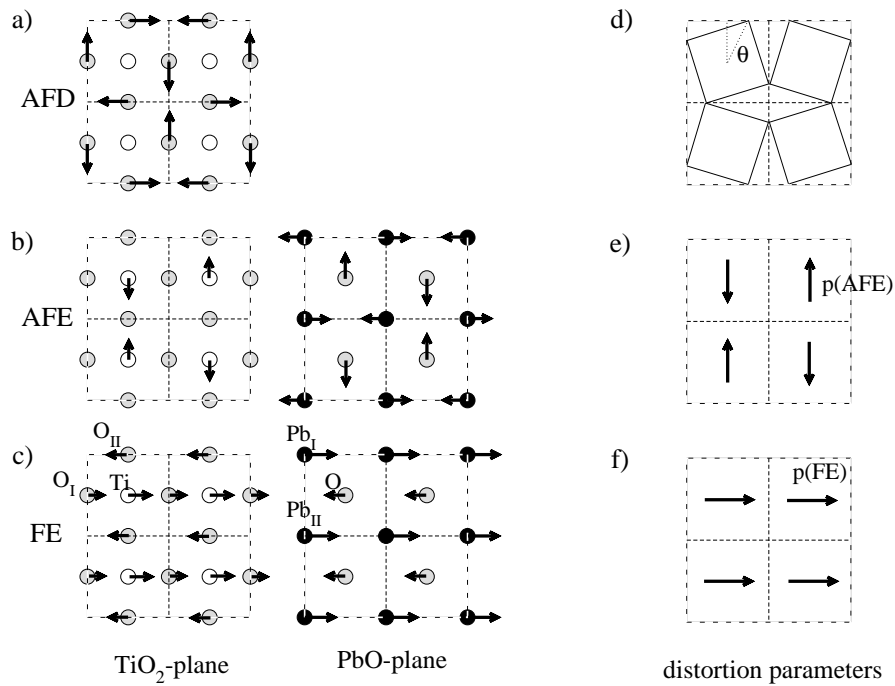

\centerline{
\hbox{
\epsfig{figure=Fig5abc.eps,height=9cm,angle=0}
\quad\quad\quad\quad
\epsfig{figure=Fig5def.eps,height=9cm,angle=0}
}}
\caption{(a), (b), and (c) atomic displacements, in the TiO$_2$ and
PbO planes, for the AFD (M$_{3+}$), AFE (M$_{5-}$), and FE ($\Gamma_{5-}$)
irreducible representations that contribute to the ground state
structure of the in-plane polarized film.  (d), (e), and (f) schematic
representation of the layer-by-layer distortion parameters (see text)
for AFD, AFE, and FE irreps, respectively. Dashed lines indicate the
(1$\times$1) unit cells.  }
\label{fig:modes}
\end{figure*}

We now consider these coupled distortions in more detail, using a 
symmetry analysis.
The c(2$\times$2) supercell can accommodate atomic distortions of the
(1$\times$1) ideal unit cell with wavevectors {\bf q}=(000) and {\bf
q}=(110)$\pi/a$ ($\overline \Gamma$ and $\overline {\rm M}$ point of
the (1$\times$1) surface BZ, respectively).  Thus, the atomic
displacements relative to the ideal perovskite structure can be
decomposed into $\overline \Gamma$- and $\overline {\rm M}$-point
irreducible representations of the reference (1$\times$1) unit cell:
$\delta_{\alpha,i}=\sum_\nu \delta_{\alpha,i}^\nu$ (where $\nu$ labels
the irreducible representations, $i$ the atoms in the c(2$\times$2)
unit cell, and $\alpha$=$x,y,z$).
This decomposition allows us to
separate the FE, AFE, and AFD contributions to the lattice distortion.
We find that the in-plane atomic displacements can be decomposed
into three irreps $M_{3+}$, $M_{5-}$, and $\Gamma_{5-}$ (or
$E_u$), which correspond to AFD, AFE, and FE distortions,
respectively. The atomic displacement patterns for the
$M_{3+}$, $M_{5-}$, and $\Gamma_{5-}$ irreducible representations are shown in
Fig.~\ref{fig:modes}(a-c).

To quantify, layer-by-layer, the strength of the AFD, AFE, and FE
distortions we proceed as follows.  The layer-by-layer
strength of the AFD contribution is described by the TiO$_4$ rotation
angle, $\theta$, in each TiO$_2$ layer (see Fig.~\ref{fig:modes}d).
For the FE ($\nu$=$\Gamma_{5-}$) and AFE ($\nu$=$M_{5-}$) contributions
we define the average polar distortion along the $\alpha$-direction,
$p_\alpha^\nu$, in each (1$\times$1) unit cell of each atomic layer (see
Fig.~\ref{fig:modes}e-f); for TiO$_2$ layers

\begin{equation}
p_{\alpha}^{\nu} =
                               \delta_{\alpha}^{\nu}({\rm Ti}) -
       \frac{[\delta_{\alpha}^{\nu}({\rm O}_{\rm I})+
             \delta_{\alpha}^{\nu}({\rm O}_{\rm II})]}  {2} , 
\end{equation}
and for PbO layers
\begin{equation}
p_{\alpha}^{\nu} = 
                   \frac{[\delta_{\alpha}^{\nu}({\rm Pb}_{\rm I}) + 
                          \delta_{\alpha}^{\nu}({\rm Pb}_{\rm II})]}{2}
                                -\delta_{\alpha}^{\nu}({\rm O}) ,
\end{equation}
where $\delta_{\alpha}^{\nu}$ are the atomic displacements in the
$\alpha=x,y$ direction relative to the ideal (1$\times$1) perovskite
structure, for each irrep $\nu$.
%  layer-by-layer average FE (AFE) polar distortion per (1$\times 1) unit cell
For a film with spontaneous polarization along $x$, the average FE
polar distortion is along $x$, $p_x^{\rm FE}$, and has the same sign in
each (1$\times$1) unit cell, while the average AFE polar distortion is
along $y$, $p_y^{\rm AFE}$, and has opposite sign in each neighboring
(1$\times$1) unit cell.

The computed values for the parameters describing the strength of FE,
AFE, and AFD distortions in each atomic layer are summarized in
Table~\ref{tab:inplane}.  
The FE distortion is enhanced at the surface and rapidly resumes the
bulk value in the inner layers. The average polar distortion in the
first PbO layer is 1.7 times larger than the corresponding quantity in
the bulk, and the enhancement is confined to the surface layer. The
same behavior for the in-plane FE distortion was found at the
(1$\times$1) PbO-terminated surface,\cite{Bernd99} where AFD distortion
is not allowed by symmetry. 
There, the average in-plane polar distortion in the first surface
layer was found to be 1.5 times larger than the corresponding quantity
in the bulk. It should be noted that in that calculation, the in-plane lattice parameters
of the slab were taken as the bulk tetragonal c and a parameters, and the comparison is to the tetragonal
bulk polarization.  In our calculation, we fixed the in-plane lattice parameters 
to match a substrate of bulk cubic PbTiO$_3$.
The proper reference bulk limit in this case is the cubic lattice 
optimized allowing a spontaneous
polarization along the (100) direction (i.e. allowing internal atomic
displacements along (100)). The resulting ``cubic polarized''
structure (C-FE) is the bulk reference limit in
Table~\ref{tab:inplane}.

The AFD oxygen-rotation is also enhanced at the surface, $\theta_2$=10.8$^o$,
and rapidly vanishes in the deeper layers. Thus, the AFD
distortion is confined to the surface octahedron layer, and is
comparable to the one found in absence of FE distortions.

Finally, there is a small AFE distortion at the PbO surface layer that 
rapidly decreases in the inner layers where it is essentially zero.

The FE and AFD distortions have both the effect
of shortening some of the Pb-O bonds. At the surface, 
the compounded effect of the coexisting large FE and AFD distortions 
results in even shorter Pb-O bonds, as short as 2.31 \AA\ (comparable to the 
shortest bondlength in PbO bulk).

%====================== FIGURE 5 ========================================
%========================================================================

\begin{table}
\caption{The average layer-by-layer ferroelectric, $p_x^{\rm FE}$, and
antiferroelectric, $p_y^{\rm AFE}$, polar distortions per (1$\times$1)
unit cell, in percentage of the bulk lattice parameter
($a_b=7.37$a.u.),  and the anti-ferrodistortive oxygen rotation,
$\theta^{\rm AFD}$, in the TiO$_2$ layers. The average spontaneous
polar distortion and the AFD angle computed for the bulk ``polarized
cubic'' structure (C-FE) and the bulk ``cubic AFD'' structure (C-AFD)
are given as a bulk reference.}
\begin {ruledtabular}
\begin{tabular} {c c c c c  c}
             & \multicolumn{2}{c}{$p_x^{\rm FE}$}
             & \multicolumn{2}{c}{$p_y^{\rm AFE}$}
             &  $\theta^{\rm AFD}$ \\
    layer    &  TiO$_2$    &  PbO  & TiO$_2$ &  PbO   &  TiO$_2$ \\
\hline
     1       &             &  11.9 &         & $-$2.4 &          \\
     2       &    4.1      &       & \ 0.5   &        & 10.8$^o$ \\
     3       &             &  6.3  &         &  \ 0.4 &          \\
     4       &    3.7      &       &$-$0.06  &        & $-$0.1$^o$ \\
     5       &             &  7.1  &         & $-$0.2 &          \\
     6       &    3.9      &       & \ \ 0.08&        & \ 0.9$^o$ \\
\hline
     bulk (C-FE)&    4.1   &  7.1  &         &        &    0$^o$  \\
     bulk (C-AFD)&   0.0   &  0.0  &         &        & \ 3.3$^o$ \\
\end{tabular}
\end{ruledtabular}
\label{tab:inplane}
\end{table}

\section{Conclusions}
\label{sec:conclusion}

Non-polar cubic PbTiO$_3$ bulk has a weak AFD instability, involving
rotations of oxygen octahedra, which is suppressed by the
FE distortion.  In this work we have studied the effect of surface
proximity upon the AFD instability at the (001) surfaces of PbTiO$_3$.
In particular we studied the surface effect upon the instability's
strength and its competition with in-plane ferroelectricity.

In the absence of FE distortions we found that the impact of the
surface upon the strength of the AFD instability is
termination-dependent.
At the PbO-termination the AFD deformation is strongly enhanced
relative to the bulk; the top surface octahedra rotate by $11^o$ about
the surface normal. The enhanced AFD distortion is confined to the
octahedron layer at the top of the surface.  We estimated that the
energy gain associated with the octahedron rotation is two orders of
magnitude larger at the surface than in the bulk.
At the TiO$_2$-termination, the AFD deformation is essentially
suppressed; the top surface octahedra (which are truncated) rotate
by an angle of $0.1^o$.
For both terminations the surface effect upon the AFD distortion is
confined to the first octahedron-layer at the surface, and below the
surface octahedra the AFD rotation recovers the small bulk-like value,
$\approx 3^o$, computed for non-polar cubic PbTiO$_3$.

For polarized films (with in-plane FE distortion) we found that FE and
AFD distortions coexist in the proximity of the PbO-terminated surface,
leading to a structural phase not found in bulk.
The FE distortion is enhanced at the surface layer as
was previously found for the (1$\times$1) films where the AFD distortions are
not allowed by symmetry. Concurrently, the AFD distortion is enhanced at
the surface as much as found in absence of FE distortions. Thus, AFD and
in-plane FE distortions are both enhanced at the PbO surface and do not
strongly interact.  The enhanced AFD deformation does not
suppress the FE distortion, nor is the AFD deformation suppressed by
the FE distortion.
%They are decoupled by symmetry at second order but can be coupled 
%at higher order.

The surface effects are confined to the first perovskite unit cell at the
surface (i.e. in the first three atomic layers), while bulk-like behavior
is recovered below it.

In summary, we have shown that in PbTiO$_3$ the strength of the
AFD instability, consisting of rotation of the oxygen octahedra, 
and its interaction with ferroelectricity are affected by the proximity 
to the surface. At the PbO termination this leads to a phase, 
not observed in the bulk,  where enhanced AFD and FE distortions coexist.
The enhancement of the AFD instability at the
PbO termination is driven by the formation of shorter PbO bonds and is
a consequence of the particular chemistry of Pb; no enhanced AFD 
distortion is found, for example, at the surface of BaTiO$_3$. We suggest
that the enhancement of the AFD instability at the PbO termination
should be common among Pb based perovskite oxides.

\section*{ACKNOWLEDGMENTS }
We thank D. Vanderbilt, G.B. Stephenson, C. Thompson, and D.D. Fong for
discussions and Krzysztof Rapcewicz for a critical reading of the
manuscript. This work was supported by Office of Naval Research through
N00014-00-1-0261. The majority of computations were performed at the
Department of Defense High Performance Computing Centers NAVO and ERDC.

%\begin{references}

%%\vfill\eject
%%\end{references}


\begin{thebibliography}{99}
\bibitem{dipole} For the case of perpendicular polarization it is
 necessary to introduce a dipole layer in the vacuum region to cancel
 the artificial field arising when a slab with a net surface dipole
 density is embedded in a periodic supercell (see
 Ref.~\onlinecite{Bengtsson99}).  This dipole correction can be used,
 as shown in Ref.~\onlinecite{Bernd01}, to study the effect of
 external electric field on the equilibrium ground state structure of
 polar films with polarization perpendicular to the surface, and
 determine their ferroelectric properties.
\bibitem{bondNote} This is our theoretical value. The experimental
value for the shortest Pb-O bonds in tetragonal PbTiO$_3$ is 2.5\AA\
from Refs.[\protect\onlinecite{PbObondRef1,PbObondRef2}].
\end{thebibliography}
\end{document}